# Infrared spectroscopy of hydration-controlled eumelanin films suggests the presence of the Zundel cation

Zakhar V. Bedran [1], Sergey S. Zhukov [1], Pavel A. Abramov [1], Ilya O. Tyurenkov [1], Boris P. Gorshunov [1], A. Bernardus Mostert [2], and Konstantin A. Motovilov [1,*]

[1] Center for Photonics and 2D Materials, Moscow Institute of Physics and Technology, Institute Lane 9, 141701, Dolgoprudny, Moscow region, Russia; info@mipt.ru

[2] Department of Chemistry, Swansea University, Singleton Park, SA2 8PP, Swansea, Wales, UK; study@swansea.ac.uk

* Correspondence: k.a.motovilov@gmail.com

**Abstract:** Eumelanin is a widespread biomacromolecule pigment in the biosphere and is intensively tested in numerous bioelectronics and energetic applications. Many of these applications depend on eumelanin's ability to conduct proton current at various levels of hydration. The origin of this behaviour is connected with a comproportionation reaction between oxidized and reduced monomer moieties and water. However, neither this reaction nor the formation of the aqueous proton species have ever been directly observed. Presented here is a hydration dependent FTIR spectroscopic study on eumelanin, which allows for the first time to track the comproportionation reaction via the gradual decrease of the carbonyl group concentration (1725 cm$^{-1}$ band) versus hydration. Furthermore, we detect two types of interfacial water (3253 cm$^{-1}$ and 3473 cm$^{-1}$ bands). Finally, the feature detected at the 3600 cm$^{-1}$ band is assigned to the formation of the Zundel cation, $H_5O_2^+$, an observation and suggestion not previously made. We suggests, due to the behaviour of the hydration dependent feature at 3600 cm$^{-1}$ that the formation of the $H_5O_2^+$ cation potentially functions as a trap for mobile proton species, partially explaining the complex hydration dependent conductivity of eumelanin.

# 1. Introduction

The melanins are one of the most diverse and widespread families of natural pigments in the biosphere [1–3]. For example, the pigments are found in plants as well as animals. In lower vertebrates, the presence of melanin in the liver and spleen suggests that it has cytoprotective functions against cytotoxic species such as activated oxygen [4–6]. More famously though, the melanins are found throughout the human body, including the eyes [7], ears [8], substantia nigra of the brain stem [9], and the skin where it acts as our main photo protectant [10].

There are two main forms of melanin, pheomelanin that is red-yellow in colour, and eumelanin that is a brown-black pigment [11]. Notably, there is a third form, termed neuromelanin, found in the brain stem and is a combination of pheo- and eumelanin [12]. Of the above categories, the most widespread melanin in the animal kingdom is eumelanin and is often considered the archetypal melanin [11]. Eumelanin is a polymeric material derived from two monomers, 5,6-dihydroxyindole (DHI) and 5,6-dihydroxyindole-2-carboxylic acid (DHICA) and their various redox states and tautomeric forms (fig. 1).

Eumelanin, the focus of our work here and referred to as melanin hereafter, has received significant attention in the last decade owing to its unique physico-chemical properties which include: broad-band optical absorption [13,14], persistent free radical signal [15–17], the ability to protect against harmful ionising radiation [18], almost 100% nonradiative conversion of light energy [13], metal ion chelation [19,20] hydration dependent conductivity [21–24], photoconductivity [21,25,26] and hydration dependent electrical switching behaviour [27–29]. Due to these aforementioned properties, melanin has been tested in numerous bioelectronic and bio-based applications, including transistor devices [30,31] biodegradable / biocompatible energetics including supercapacitors [32–35] and elsewhere [36–39]. Even though melanin is not a monolithic material and the ratio between DHI and DHICA-based moieties in the polymer in the dry state depends on the particular synthetic procedure used / origin of material [40], it is clear that there is a significant effect on melanin's electrical properties in the presence of water, where a redox equilibrium (fig. 2) generates a semiquinone anion and charged proton species, which leads to an enhanced electrical conductivity of the material [21,24,41,42]. Since the conductivity is a crucial parameter for device applications, attempts have been made to modify it via non-hydration means such as with metal ion chelation [31] and sulfonation [43].

However, even though the aforementioned redox reaction (comproportionation) is a source of proton charge concentration for conductivity, there has been little to no focus on the mobile nature of these protonic charges. It is not clear at all whether the proton mobility remains constant or whether it changes as a solid state melanin sample is hydrated. Furthermore, it is also not clear whether the ubiquitous assumptions of the $H_3O^+$ cation as the main form and associated Grötthus mechanism [44] are correct [45].

To address some of the shortcomings in the literature regarding the nature of the protonic charges, we report here a Fourier Transform Infrared (FTIR) study of synthetic melanin films. The films were probed under thoroughly controlled, various levels of hydration at room temperature (25ºC) in order to observe the behaviour of the melanin as well as the water. FTIR is an excellent tool to investigate the potential formation of different hydrated protonic charges, since each species will have a different vibrational signature.

Vibrational spectroscopy including Raman scattering and FTIR techniques, excepting neutron inelastic scattering (INS), has been used to characterize melanin materials numerous times [4,15,46–60]. Particular efforts were aimed to clarify origination of spectral peculiarities of melanin from definite monomer units [54,61] dimers [51,62] and tetramers [51,60]. Only a few studies were performed with precise attention towards the level of hydration of the material and the corresponding consequences for spectral features [46,52,63]. However, even in these latter cases the interpretation of the data was based only on transformations of melanin monomer units, swelling and shrinking of internal distances between substructures caused by the hydration as well as changing ratios of strongly and weakly bounded water. The aforementioned problem of the formation of various hydrated forms of proton charges, remains completely untouched in melanin vibrational spectroscopy.

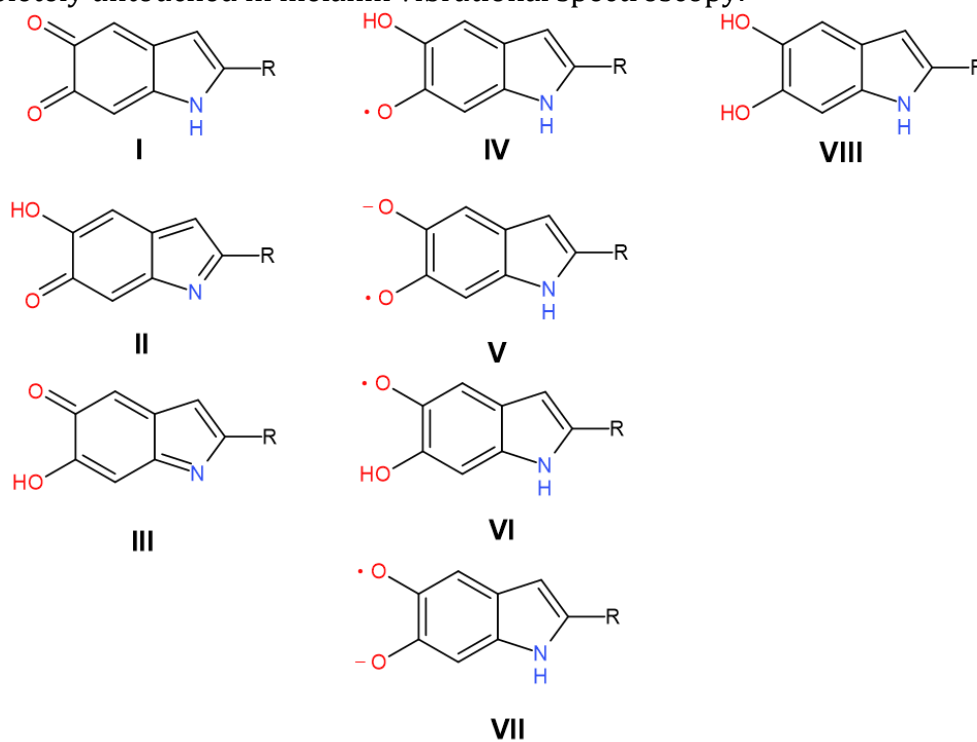

**Figure 1.** Redox states and tautomeric forms of monomers within the eumelanin chain. R is H (for DHI) and COOH (for DHICA). Left column includes fully oxidized forms: I - quinone, II - quinone methide, III - quinone imine. Central column contains semireduced radical forms: IV, VI - protonated semiquinones, V, VII - deprotonated radicals. Right column contains a single fully reduced form - VIII - hydroquinone.

Thus, unlike previous studies, this current work is aimed to reveal potential transformations of basic protonic species (such as the hydronium cation and others) within melanin, a likely occurrence given the course of hydration and the active comproportionation reaction (fig. 2). We go on to juxtapose our observations with already published data on hydration dependent AC and DC conductivity [15,22–24,42,64], muon spin resonance (muSR) scattering [21] and electron paramagnetic resonance (EPR) [64–66].

In relation to the conductivity of melanin, the current explanation of synthetic melanin's conductivity [21,23,24,31], muon spin relaxation (μSR) [21] and solid state magnetic properties [64–66] with respect to the level of hydration is based on the comproportionation reaction fig. 2. In essence, water addition leads to the formation of the semiquinone radical and proton generation via equilibrium chemistry.

Notwithstanding almost one decade of studies of melanin as a system with probable water-dependent comproportionation process, as far as we know, no one has reported observations of this reaction, in the solid state, as a change of chemical moieties, i.e. change in chemical composition induced by water. The best evidence of the solid state comproportionation was manifested as a strikingly similar complex hydration-dependent increase of conductivity $\sigma_{DC}$ as well as $\lambda$ and $\Delta$ components of muon spin relaxation, responsible for relaxation rates of para- and diamagnetic muons respectively [21] as well as a kinetic isotope study in which heavy water was employed [64]. Solid state EPR results gave more indirect arguments for comproportionation [64–66] because of the inability to separate signals of protonated and deprotonated semiquinone from each other and general increase of radicals concentration in dried melanin [15,17]. Given the above situation, in addition to our investigation of the nature of the water and proton species in melanin, we will also investigate the potential chemical reaction within the solid-state melanin as it pertains to the comproportionation reaction.

There are a couple of additional observations to be made at this point for this study. First, hydration at room temperature, without direct contact with liquid water, may give up to 18% weight gain in water content in melanin depending on the relative humidity [67]. Secondly, the vibrational spectroscopy of water and different proton species is well studied in inorganic solid state ionics and ionic liquids [68–71]. The fingerprint frequencies for the stable cations $H_3O^+$, $H_5O_2^+$ and heavier proton species in polar environments are well known in FTIR, Raman scattering [68] and INS [71–73]. The unambiguity of assigning the corresponding vibrational modes to these ions is also supported by DFT modelling modeling and X-ray structural analysis data [74,75]. The knowledge bank above makes interpreting our data far more straightforward. This is especially pertinent given that melanin has not been successfully computationally modelled, due to its chemical heterogeneity making it computationally expensive (see for example a recent study where only up to tetramers were modelled [76–78]). Thirdly and finally, the gradual hydration-dependent transitions between more and less mobile aqueous proton cations observed by means of quasi-elastic neutron scattering (QENS) [79] and various NMR techniques [80–83] in solid state proton conductors offer another avenue for understanding the case of hydrated melanins. We will apply this knowledge to expound on our observations below and explain peculiarities of hydration dependent behaviour of melanin polymer physical properties.

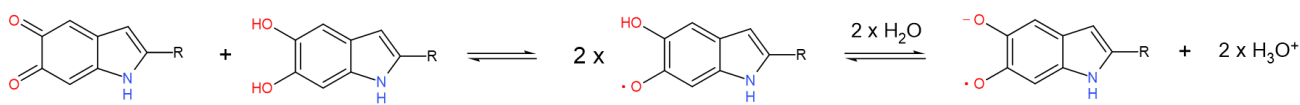

**Figure 2.** The comproportionation reaction, where the oxidized form (quinone I or quinone methide II or quinone imine III) and the reduced form (quinol VII) of moieties leads to the formation of the intermediate oxidation form, the radical semiquinone (IV or VI). In the solid state, hydration leads to deprotonation of semiquinone to form semiquinone anion (V or VII) and mobile proton species, traditionally signified on schemes by hydronium cations $H_3O^+$.

## 2. Materials and Methods

*2.1. Melanin synthesis*

Melanin was synthesized following a standard literature procedure [19] utilising as the initial starting material d,l-dopa (Sigma-Aldrich). D,L-dopa was dissolved in deionized water, subsequently adjusted to pH 8 using $NH_3$ (28%). Air was then bubbled through the solution while being stirred for 3 days. During the 3 days synthesis the pH would naturally decrease due to the $NH_3$ evaporation, which would necessitate the need to add ammonia periodically to bring the pH back to 8. By keeping the pH at a maximum of 8 and letting it decrease naturally ensured that ring fission of the indolquinone moieties are kept to a minimum and hence the synthesised melanin is a biomimetic material [84]. The solution was then brought to pH 2 using HCl (32%) to precipitate the pigment. The solution was then filtered and washed multiple times with deionized water and dried. We made 4 batches and then homogenised them to make a representative material.

*2.2. Melanin characterisation*

The samples were confirmed as melanin via UV-Vis absorbance spectroscopy, electron paramagnetic resonance (EPR) and x-ray photoelectron spectroscopy (XPS).

For the UV-Vis absorbance spectroscopy a small amount of melanin powder was dissolved in a pH 8 deionised water solution (adjusted with $NH_3$). Spectra were then obtained utilising a Perkin Elmer PDA UV/Vis Lambda 265, using the wavelength monitoring functionality, obtaining a wavelength range from 350 nm to 900 nm at 1 nm wavelength intervals. Example data can be seen in fig. 3A.

For the EPR measurement, a powder sample of the melanin was measured using a Bruker EMX Micro X, CW-EPR spectrometer with an E4104 X-band cavity at a microwave power of 0.87 mW and room temperature. The spectra were taken at a modulation frequency of 100 kHz, modulation amplitude of 1 Gauss with a scan width 60 Gauss. An example spectrum can be seen in fig. 3B.

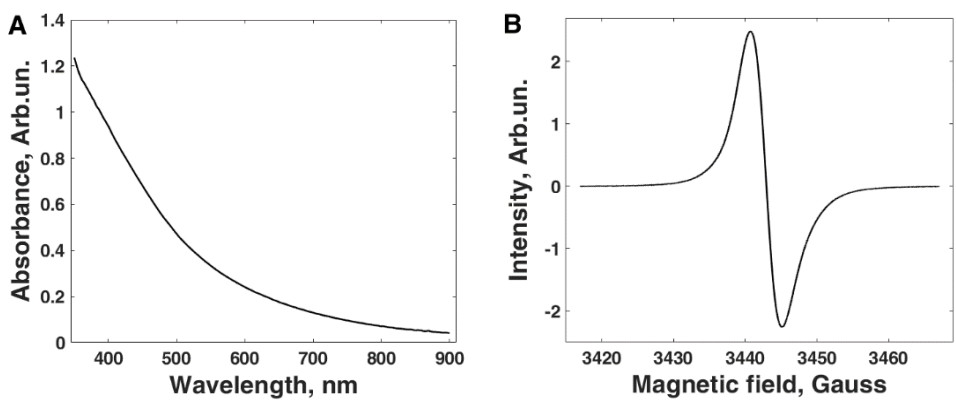

**Figure 3. A**) An example UV-Vis absorbance spectrum obtained for the melanin sample. The curve shows a simple decaying exponential as expected for the material. **B**) An example CW-EPR X-band spectrum obtained for the sample.

Obtaining an elemental analysis for melanin can be reliably probed via XPS. For the XPS a wide-scan survey spectrum was performed on pressed pellets of the powder utilising a Kratos Axis Supra using a 225 W AlKα X-rays with an emission current of 15 mA and equipped with a quartz crystal monochromator with a 500 mm Rowland circle. Spectra were collected with a pass energy of 40 eV, with the hybrid lens setting, 0.1 eV step size, 1 s dwell time for electron counting at each step. The integral Kratos charge neutralizer was used as an electron source to eliminate differential charging.

*2.3. IR-measurments*

Transmission spectra in the middle infrared region (MIR) were obtained by utilising a Bruker Vertex V80 FTIR spectrometer with Hyperion 2000 microscope console, equipped with IR objective x36 (NA 0.5, working distance 10 mm) and 50 micrometer aperture.

Humidity control was achieved by means of a custom hygrostatic system shown in fig. 4A & B. The container (1) with a saturated salt solution was equipped with a small fan inside responsible for creation of steady moisture distribution throughout the volume. The modified APS 300 pump (Tetra) (2) was used to circulate moisturized air through the system. The saturated salt solutions used to achieve humidity with their corresponding relative humidities (RH) can be seen in table 1. The salts were obtained from Sigma-Alrich (LiCl, KCl), Rushim (MgCl$_2$, Na$_2$Cr$_2$O$_7$, NaCl) and used as is.

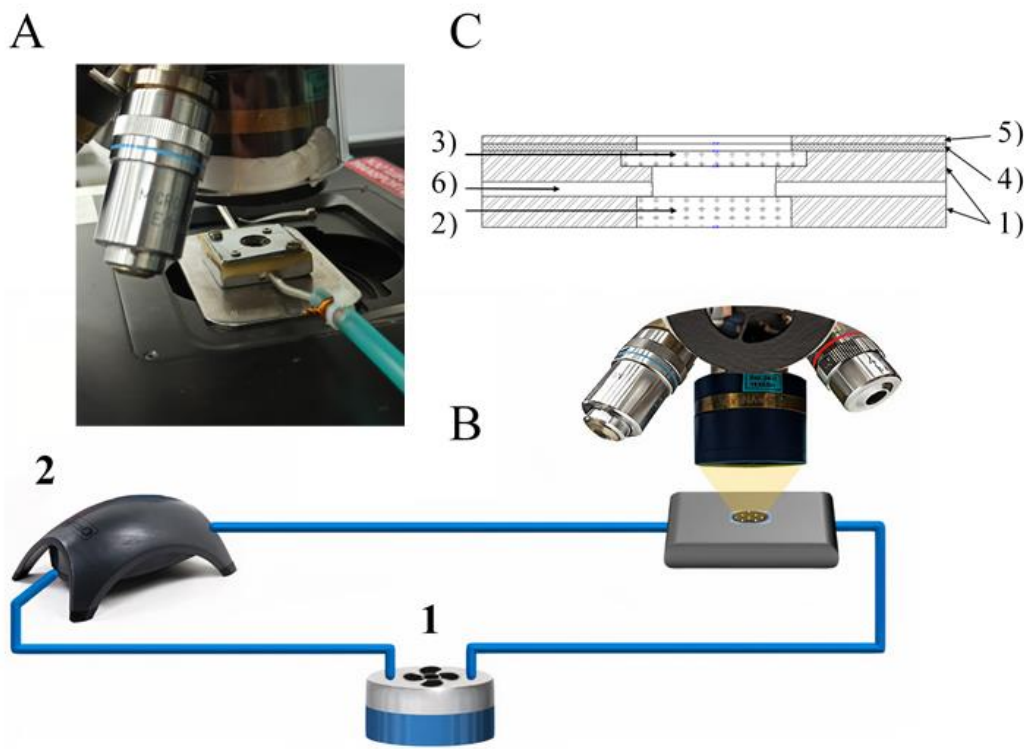

**Figure 4.** An overall scheme of the experimental setup. In inset **A**) a photo shows the real life setup. Inset **B**) depicts the sample (in the microscope setup) connected to a container with a saturation salt solution (**1**) equipped with a fan, which ensures a steady moisture distribution in the container. The recirculation of the moisturized air through the sample and the whole system done by the pump (**2**). The inset **C**) contains the drawing of the hygrocell described in text.

**Table 1.** A list of the saturated salt solutions and their corresponding relative humidities (RH) at room temperature.

| Salt | LiCl | MgCl$_2$ | Na$_2$Cr$_2$O$_7$ | NaCl | KCl |
|---|---|---|---|---|---|
| RH at 25°C, % | 11 | 33 | 54 | 75 | 84 |

The minimal value of relative humidity was achieved by venting for 15 hours of hygrostatic cell with gaseous nitrogen with 1.4% RH. We estimated moisture content in nitrogen by optical transmission measurement with a good resolution (0.2 cm$^{-1}$) in a Bruker Vertex V80 sample compartment. Water content was calculated via analysis of line spectral strength by means of HITRAN [85] open spectroscopic database and load_hitran MATLAB function [86].

The key part of hygrostatic system is the hygrocell depicted on fig. 4C. It allows us to measure transmission spectra in MIR while changing the hydration state *in-situ*. The body (1) of the cell was made of low-carbon structural steel. The window (2) was made of CaF$_2$. It was glued to the body with epoxy resin. The examined samples were sprayed on the bottom of the replaceable window (3) also made of CaF$_2$. The window (3) was sealed with a rubber gasket (4) and fixed by a steel cover (5). Moisturized air was supplied into the hygrocell through the central 1 mm wide channel (6).

For spraying we prepared the solution in 20% aqueous ammonia of synthetic eumelanin with concentration of 45 mg per 1 ml. The solution was stirred for 1 hour and then ultrasonicated for 1 hour. Then the solution was sprayed via an airbrush from a distance of 10-15 cm onto a $CaF_2$ substrate previously heated at 50°C and cleaned with isopropyl alcohol. We utilised a nozzle with a diameter of 0.5 mm, the air pressure in the airbrush was 1 atmosphere. The thickness of the sample films was estimated by means of SOLVER NEXT Scanning Probe Microscope (NT MDT, Russia) and was found to be around 1.5 micrometers.

To achieve equilibrium between sample and the moisturized atmosphere the pumping procedure was performed continuously for 1 hour at the desired level of humidity. We then measured the MIR spectra twice with a 10 minutes delay. We draw the conclusion that the equilibrium had been achieved if two sequential test spectra were the same.

*2.3. Data processing*

The measured infrared spectrum was deconvoluted into an appropriate set of Gaussian and Lorentzian spectral lines following the procedure below using Fityk data processing software. At the pre-processing stage we subtracted the baseline caused by the scattering and reflection from the all absorption spectra. As one of the most powerful tools for detection of overlapping peaks we used next the second derivative peak finding procedure [87]. For data processing we used the positions of local maximum at the second derivative graph as an initial guess for the Gaussian line central frequencies for 1.4% RH. We estimated corresponding Gaussian peak widths and heights manually and then utilised a Levenberg-Marquardt least square difference minimization algorithm (L-M LSDM) to obtain a refined result for all parameters. Then we found that a set of Gaussian lines fits well with the experimental data except the mode with 1718 $cm^{-1}$ peak position. Due to its wide slope we decided to use the Lorentzian line shape instead of the Gaussian one for this particular line. The derived parameters were used as an initial guess for processing of data obtained for higher humidities. Since we did not observe reliable frequency shifts during hydration in the second derivative graph (ESI fig. 1), we fixed the frequency positions for all modes. This approach made data processing dependent on the model and on the starting humidity point. To overcome this drawback we utilised the following procedure. We unfixed the first frequency for particular RH value, then applied the L-M LSDM algorithm, recorded the derived frequency value, returned the initial number for it and repeated it for all frequencies. Doing this for each value of humidity we have received the variation of the central frequencies with RH change and, thus, we were able to estimate the mean value for each frequency with RH variation. These mean values were used for further data processing. Repeating the described frequency variation procedure we were able to double-check stability of the chosen frequency set and to estimate the error values for frequency calculation. The results are listed in Table 3 and ESI Table 1.

Thereafter, we found a mutual dependence between the Gaussian parameters of pairs of modes with 2590 $cm^{-1}$/ 2769 $cm^{-1}$ and 2849 $cm^{-1}$ / 2932 $cm^{-1}$ peak positions on the second derivative curve. To improve fitting quality we united these four modes into two peaks with Gaussian central frequencies 2600 $cm^{-1}$ and 2853 $cm^{-1}$. Moreover, such treatment was supported by similar assignments for united modes in cited literature [63]. Also, we observed almost static behavior of the absorption line with 3368 $cm^{-1}$ central frequency. Considering this, we fixed all the parameters of this latter line to enhance fitting quality. The evolutions due to hydration of the relative change in line strength of the observed excitations are listed on the text below and in ESI.

*2.4. Reversibility check*

To check the reversibility of the observed line strength dynamics we performed the following procedure. We reproduced the initial experimental approach described above in the section 2.3, but added also the backward sequence of relative humidities. Thus, the full sequence of RH in this case was: 1.4% → 11% → 33% → 54% → 75% → 84% → 75% → 54% → 33% → 11% → 1.4%. The independence of the absorbance second derivative on the moisture content was checked. We demonstrate it in ESI Figure 5.

For data processing we used previously obtained set of frequencies. As one can see from the ESI Figure 4, it worked well for almost the all lines except the pair 2849 cm$^{-1}$ / 2932 cm$^{-1}$. In first experiment without backward series of humidities we merged lines into one with frequency 2600 cm$^{-1}$, but in the new dataset the line 2936 cm$^{-1}$ was fitted better. Taking into account the same assignment for these lines we accept this change. The existence of the 3850 cm$^{-1}$ mode in the new data is questionable since we observed much smaller absorbance at this frequency. The baseline correction in the case of the new dataset should be done by a straight line with a slight slope instead of a horizontal line as it was performed previously. The change in baseline extraction procedure makes the interpretation of the highest frequency small mode more model-dependent. For the other lines only the slight frequency shifts have been observed as it can be seen from the legends on ESI Figure 4.

## 3. Results

The characterisation results of our sample utilising UV-Vis, EPR and XPS can be seen in fig. 3 and Table 2. The UV-Vis data in fig. 3a show a featureless, exponential decay spectra is as expected for melanin [88]. The EPR spectra shown in fig. 3B has an apparent isotropic g factor, calibrated against a DPPH standard, of 2.0036, which is as reported elsewhere for solid melanin synthesized in the same manner [43]. Finally, as shown in Table 2, the atomic composition (atomic concentration in at% and atomic ratio) of the sample can be seen and is compatible with that of a synthetic melanin. We note that elemental surface scans of melanins are representative of the bulk as previously demonstrated [89]. Overall, the sample is a synthetic sample of eumelanin.

**Table 2.** The atomic composition (atomic concentration%) and atomic ratios determined from pressed powder pellets of the synthetic melanin sample. For comparison the expected ratios for the monomer building blocks DHI and DHICA are shown.

| Sample | C (at%) | O (at%) | N (at%) | C/N | O/N | C/O |
|---|---|---|---|---|---|---|
| DHI - Expected | 72.7 | 18.2 | 9.1 | 8 | 2 | 4 |
| DHICA - expected | 64.3 | 29.6 | 7.1 | 9 | 4 | 2.2 |
| Sample | 69.5±0.3 | 21.2±0.4 | 9.2±0.2 | 7.6 | 2.3 | 3.3 |

FTIR absorbance of the melanin film equilibrated with correspondingly moisturized atmosphere in the wavelengths range from 1000 cm$^{-1}$ ending on 4000 cm$^{-1}$ is shown in fig. 5. The fine structure in intervals 1200-1500 cm$^{-1}$, 2700-2800 cm$^{-1}$ and above 3700 cm$^{-1}$ was found to be noise as it was not reproduced at constant RH value. Also shown are the deconvoluted spectra following the well-known second derivative line detection procedure as described in the Materials and Methods.

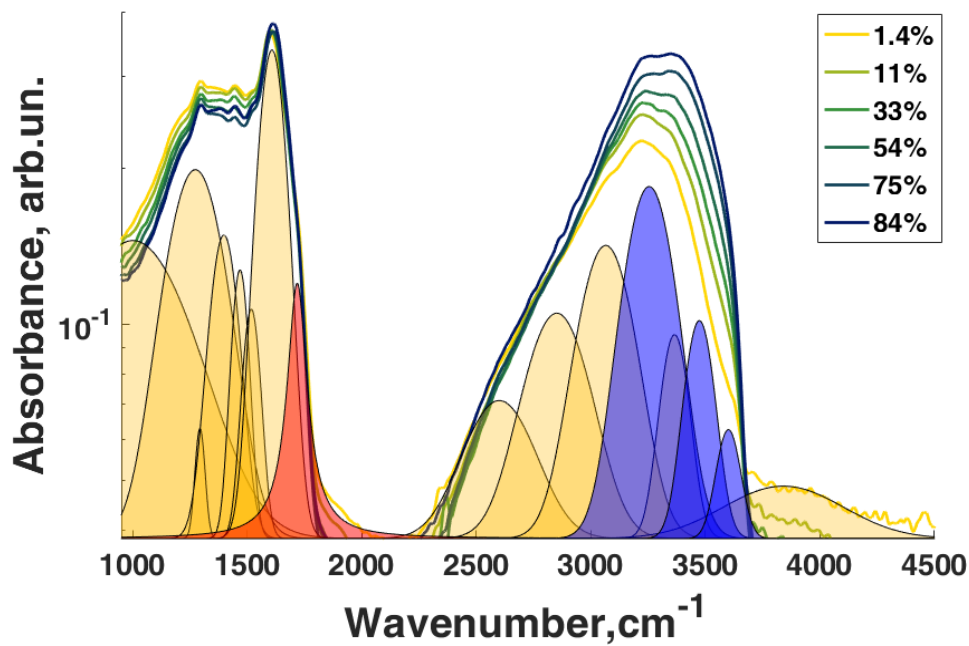

**Figure 5.** The solid lines are the measured infrared spectra of the synthetic melanin thin film at 25ºC for various hydration levels from 1.4% RH to 84% RH (see legend). The coloured peak areas are the result of the 1.4% RH spectra deconvolution into an appropriate set of well-known peaks. The red area is associated with the C=O vibration, the three blue peaks are assigned to water vibrations and the remainder, yellow peaks referred to the excitations of the melanin.

Since the peaks on the curves of the second derivative of absorption had no dependence on moisture content (ESI fig. 1), we concluded that the absorption peak frequencies remain fixed too. Peak positions and their assignments are listed below in table 3 and in ESI table 1.

**Table 3.** Positions of key observed peaks and their assignments. The remained is listed in the ESI table 1.

| Line peak position | Assignment |
|---|---|
| 1725 ± 1 | $\nu(C=O)$ in ketone or carboxylic acid [58] |
| 3253 ± 3 | Stretching vibration of the Ice-like hydrogen-bonded water at melanin particles interface [90] |
| 3473 ± 1 | Stretching vibration of the non-interface liquid-like water [91] |
| 3600 ± 2 | $H_5O_2^+$ cation fingerprint [92] |

In further analysis we wanted to derive concentration of infrared radiation absorber following the Beer-Lambert-Bouger law, which states that absorption coefficient at a certain frequency is a product of optical path length, absorber concentration and molar extinction coefficient at that frequency. However, the peaks from different absorption species are often overlapping and thus making the direct calculations of concentration impossible. To overcome this problem one needs first to deconvolute spectra into an appropriate set of well-known line shapes and then to analyze them separately. To reduce the noise effect on calculations the integral of the absorption coefficient over the mode is usually used instead of absorption coefficient at a certain frequency itself. The derived value is called the line strength (L) and is measured in cm$^{-1}$. Thus, the corrected form of the Beer-Lambert-Bouger law for the line strength will be as follows: L = x × c × ε, where L is the line strength, x is optical path, c is the sought concentration of absorption species and ε is the integral molar extinction coefficient. Determination of the ε is possible for simple chemical objects. However, in the case of our material we don't know ε of the different absorption modes. Therefore, direct estimation of the absolute values of concentrations are impossible. One result is that we are unable to potentially estimate the relative oxidative state of the material (i.e. ratio of quinone/tautomer vs hydroxyindoles), though along with Serpentini *et al.* [93] we believe our sample to be mainly in oxidised form due to the oxidative nature of the synthesis. However, we can analyze the relative change of line strength by normalizing it (i.e. the peak area) to the average line strength value obtained across all levels of relative humidity. As a result, we obtain the value directly proportional to the relative change in concentration of absorbing species. The uncertainties were estimated from the L-M LSDM fitting algorithm. The uncertainty in the raw data due to the noise were neglected as we found it much less than the uncertainty due to the fitting (0.2% for noise and >1% for fitting).

As a result of the described data processing procedure, we have obtained the frequency set, which represented the IR spectra of melanin. Then in order to check the reversibility of the obtained data we reproduced the same procedure, but added also the set of measurements performed under the reversed order of relative humidities, from wet 84% to dry 1,4% RH. To process the obtained data, we utilised the same frequency set which was obtained during the processing of the experiments with only from dry to wet sequence of applied humidities. The check for reversibility of the corresponding line strengths demonstrated that it does take place with slight hysteresis observed, however, previously in the study by Bridelli and Crippa [52].

The first key feature presented is the dynamics of the 1725 cm$^{-1}$, C=O carbonyl mode in fig. 6. The most likely sources of carbonyl groups in the material are oxidized forms of monomers I-III and carboxylic acid groups (fig. 1). As such, we assign this line among the discussed melanin units [58]. As we can see from fig. 6, the line strength of 1725 cm$^{-1}$ mode decreases with increasing RH. This is consistent with the oxidation transformations due to comproportionation reaction (fig. 2), namely, the lowering of the C=O bond concentrations due to consumption of the quinones.

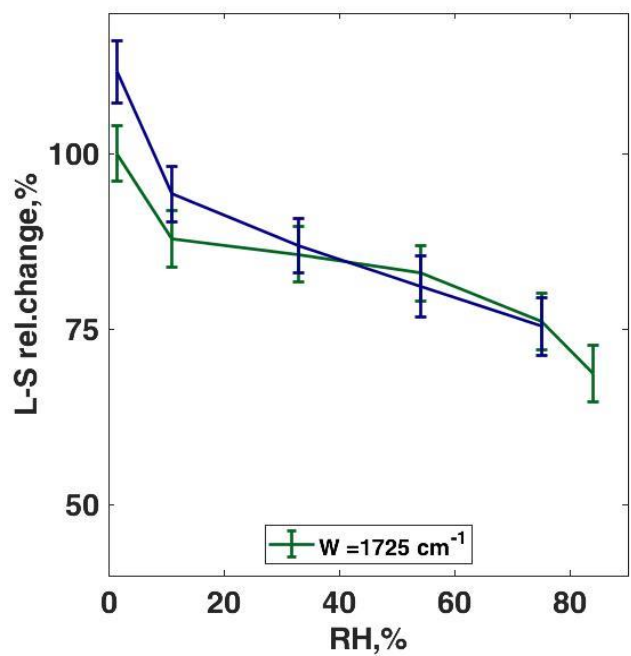

**Figure 6.** The observed hydration level evolution of the relative change in line strength of the mode with 1725 cm$^{-1}$ central frequency. On the legend, W refers to the central wavenumber of the mode in cm$^{-1}$. The error bars were estimated from the L-M LSDM algorithm. The green lines referred to the increase of RH during the measurements and the blue ones referred to the measurements with decrease of RH.

In the canvas of our work the most important excitations are intramolecular vibrational modes of water and aqueous proton species. We can clearly see the strong absorption line with 3253 cm$^{-1}$ central frequency. The line strength dependence on humidity is shown on fig. 6A. Analyzing the absorption spectra of ice by means of Gaussian deconvolution one could see the strong line with central frequency at 3240 cm$^{-1}$ [90].

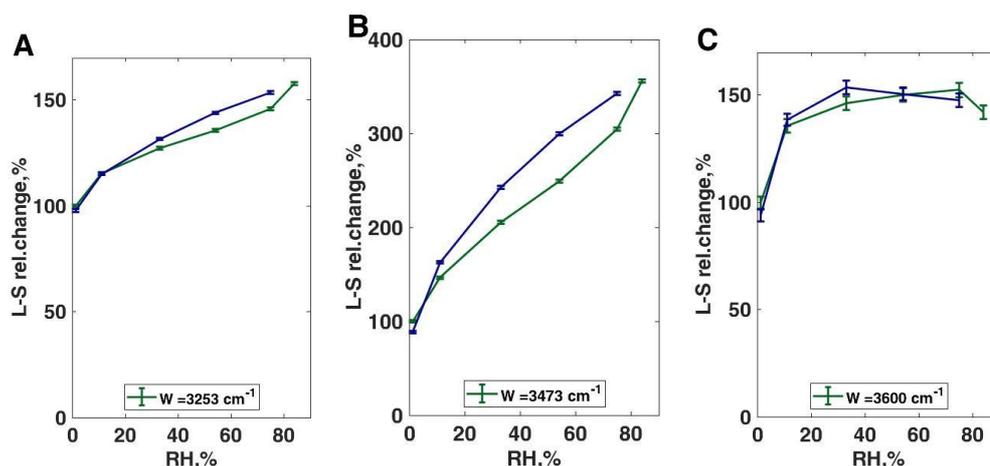

**Figure 7.** The observed evolution with hydration of the relative change in line strength of the modes with **A**) 3253 cm$^{-1}$, **B**) 3473 cm$^{-1}$ and **C**) 3600 cm$^{-1}$ central frequency. On the legend, W refers to the central wavenumber of the mode in cm$^{-1}$. The error bars were estimated from the L-M LSDM algorithm. The green lines referred to the increase of RH during the measurements and the blue ones referred to the measurements with decrease of RH.

For H$_2$O molecules surrounding melanin particles is similar to that in water ice because of a large number of hydrogen donors and acceptors on the melanin surface. Considering the listed reasons and in accordance with work [52] we attribute 3254 cm$^{-1}$ band to the stretching vibrations of water with saturated hydrogen bonds at the interface of melanin particles. The slight frequency shift of the observed excitation from the aforesaid in ice could be caused by the different d(O-O) distance between water molecules on the melanin particle surface in comparison with ice. Another explanation is a chemical shift due to the different bond strength in water-melanin interaction in comparison with water-water bonds in ice.

The next observed excitation related to water is at 3473 cm$^{-1}$. The line strength of this mode is rapidly increasing with the RH growth fig. 7B. Our interpretation of this excitation comes from analysis of the liquid water infrared spectra [91]. Following the deconvolution of absorption spectra in the range 2660 cm$^{-1}$ - 4000 cm$^{-1}$ we find two absorption modes with central frequencies around 3300 cm$^{-1}$ and 3464 cm$^{-1}$. The origin of these excitations with respect to microscopic arrangement of water molecules have been studied for decades. And according to one of the latest reviews on this topic [94] there is no clear answer up until now. However, we can use a simple expression of the vibrational frequency dependence on the distance between nearest oxygens [95,96]. It leads to assignment of the observed 3473 cm$^{-1}$ band to $H_2O$ molecules connected with other water molecules in a more bulky environment (compared to water yielding band 3254 cm$^{-1}$). The absence of the 3300 cm$^{-1}$ can be caused by insufficient distance between melanin particles for formation of long inter water bonds.

The next contribution of water into the IR spectra of melanin which attracted our attention was the band at 3600 cm$^{-1}$ (fig. 7C). Its line strength initially increases, then flattens and even decreases with the increase of water content. We regard this band as a fingerprint of $H_5O_2^+$ cation formation. According to [92] the increase of the ratio $[H_5O_2^+]/[H_3O^+]$ in the studied triflic acid:water mixtures in acetonitrile leads to an increase of the frequency of the corresponding O-H stretching vibrations, thus, attributing the highest vibrational frequencies to Zundel ion. Organic hydroxyls (in condensed states) do not manifest eigenfrequencies of these values.

## 4. Discussion

We begin by noting that, since we are unable to distinguish between the different hydroxyl groups of the various oxidised moiety forms (fig. 1) such as the quinone tautomers (present in large quantities [97]), semiquinones and hydroxyindoles, or amine/imine groups, we cannot use the change in behaviour of these peaks as evidence for, or against the solid state comproportionation reaction. Instead we focus on the carbonyl group, since a simple observation of water-induced reversible decrease of carbonyl groups concentration (fig. 2). This should be a clear and obvious consequence of comproportionation in the system since carbonyls are absent in reduced and semi-oxidized moieties. Hence, fig. 6 is to the best of our knowledge, the first study to demonstrate a reversible hydration-dependent decrease of carbonyl groups in melanin in full accordance with comproportionation reaction. As it was previously observed for $\sigma_{DC}$, $\lambda$, $\Delta$ and many other properties of melanin, the dependence of the line strength corresponding to carbonyl groups in fig. 6 is far from a simple linear dependence and contains several stages (see also fig. 7). Clarification of the nature of these stages and the origin of transitions between them is likely not only based upon the comproportionation of the melanin monomer moieties. In the following discussion we offer an explanation of the multiphase structures seen in our work and connect it to the various hydration-dependent melanin properties, which we base on the stability of aqueous proton species at different levels of hydration. The effect of hydration on proton species stability have been studied in detail in a range of inorganic proton conductors that gives us a good background for potentially understanding the IR data and results of previous melanin studies.

We first turn to the two other water-associated absorptions in fig. 7A and 7B (3253 cm$^{-1}$ and 3473 cm$^{-1}$), which shows a gradual increase with hydration across the RH range. Being responsible for tightly bound water at the interface of melanin particles the 3253 cm$^{-1}$ band gives a very smooth increase, unlike the band 3473 cm$^{-1}$. Since our starting point of hydration was 1.4% RH, some water should have been present in the system at this point, but mostly in interfacial form. Together with the finiteness of the surface of the melanin particles, the smooth linear increase for the 3254 cm$^{-1}$ feature is explainable. However, the space between melanin particles allows for greater water adsorption of less tightly bound water, which is signified by the steeper gradient behavior of the 3473 cm$^{-1}$ band, which has a shape similar to that of an adsorption isotherm.

Turning now to fig. 7C, which we indicated above is a feature for the $H_5O_2^+$ (Zundel) ion. This assignment requires an additional, deeper explanation Currently, it is generally accepted that for the stable formation and predominance of the $H_3O^+$ in preference to other proton species require special conditions [68]. The most critical of them are the presence of strong acidic groups in a structure with the ratio between these groups and water molecules being nearly 1:1. Such groups are absent in melanin. Carboxyl groups of DHICA monomer units have a pKa of 4.25 [98]. For the DHI semiquinone the pKa value is estimated to be 6.8 [99]. In the excess of water the formation of energetically more favorable proton structures takes place instead of $H_3O^+$ [68]. In the work [100] authors analyzed the enthalpy and entropy changes in the course of the reactions $H^+(H_2O)_{n-1} + H_2O = H_+(H_2O)_n$ in the gaseous phase. They found that formation of $H_5O_{2+}$ cation from hydronium cation and water molecule gives 31.6 kcal mol$^{-1}$ to the enthalpy. This value corresponds to ~ 1.37 eV making the H-bond in gaseous $H_5O_{2+}$ relatively strong. For comparison, the typical values of H-bond energy in peptides cover ranges from 0.5 to 6 kcal mol-1 depending on the type of interacting groups and water concentration in each case [101]. Addition of the next water molecules with formation of $H_7O_{3+}$ cation gives only 19.5 kcal mol-1. Further increases in n leads to a monotonous decrease in the enthalpy. Furthermore, the stable existence of $H_5O_2^+$ in solid state proton conductors, mostly hydrates of strong acids, with definite levels of water content has been known for decades [73,74,80,102,103]. Published XRD data demonstrates an almost equidistant position of the hydrated proton in $H_5O_{2+}$. The distance between two oxygens in this cation is significantly shorter (<2.45 Å) than in common systems with hydrogen bonds [75]. Therefore $H_5O_{2+}$ cation should be regarded as a stable and thermodynamically favourable form of proton aggregation at acidic pH in water-containing systems (such as hydrated melanin) at room temperature and below. As such, we believe fig. 7C is due to $H_5O_2^+$-associated absorption.

Given the assignments from, and discussion of, the literature above, we believe that fig. 7C is good evidence for due to $H_5O_2^+$-associated absorption in melanin. However, we stress here that this assignment is not proof of the $H_5O_2^+$, since demonstrating the presence of a particular proton cation requires more investigation. In the framework of this study, we limited ourselves to FTIR spectroscopy. To obtain more reliable evidence of $H_5O_2^+$ presence in melanin one needs to follow a 12-tunstophosphoric acid (TPA, see ESI) and other inorganic proton superionics research path. Experiments should include Raman spectroscopy and try to find features of $H_5O_2^+$ ion such as the intensive band near 500 cm$^{-1}$. Inelastic neutron scattering, a sister technique to FTIR and Raman will also be needed to fully flesh out the details. Unfortunately, melanin cannot be crystallized and its polyradical nature makes it difficult to study by means of NMR techniques. Nevertheless, other physical methods can also assist. Since the formation of $H_5O_2^+$ cation with relatively low dipole moment (compared to $H_2O$ or $H_3O^+$) should decrease the general dielectric contribution Δε of the well-known wide Debye relaxation [104]. Thus, the dielectric spectroscopic response of melanin in the GHz-THz range should be also examined.

With the above in mind, it is interesting to speculate what the consequences are, if indeed the Zundel ion is responsible for the behaviour in fig. 7C, on melanin's conductivity behavior. Inspecting fig. 7C in more detail, the bell-shaped curve (which peaks at RH ~54%) is not fully understood, though the literature may provide some insight. First, we draw the attention of the reader to the maximum point (~55% RH). This peak is quite close to reported hydration dependent conductivity data of melanin thin films [42] (the morphology we examine in this work). Indeed, this work by Sheliankina et al. shows a hydration dependent conductivity that follows the same trend as depicted in fig. 7C, i.e. a large increase at low RH, a peak at around 50% RH, and then a slight decrease and apparent levelling off. What we propose is that at low hydration levels the conductivity increases via the comproportionation reaction as has been discussed above. This is accompanied by the steady formation of Zundel ions in the material, but which does not impact the conductivity materially. However, the conductivity reaches a peak and then starts decreasing just around or before the peak in the Zundel ion formation occurs, essentially implying that further formation of the Zundel ion is the cause of the decrease and levelling of the conductivity. This latter suggestion is not that unusual. Multiple studies performed on solid state inorganic acids gave generally the same image for dependence of $H_5O_2^+$ cation concentration on hydration level. Moreover, according to those studies the increase of $H_5O_2^+$ content always leads to decrease of mobile proton species concentration in the system and, consequently, to decrease of the conductivity (see ESI for a discussion on example materials, see [72,79–81,83,105–107]). Thus, in the framework of our logic it means that the reaction $H_3O^+ + H_2O = H_5O_2^+$ functions as a trap for mobile protons species within melanin.

However, this change in mobility argument has to be weighed in light of previous work. In melanin the mobility of muons (a proxy for proton mobility) was demonstrated to be completely independent from the level of hydration [21]. It means that all the observed hydration-dependent changes of $\sigma_{DC}$ should be reduced to the changes of concentration of mobile proton species, but not the changes of proton mobility mechanism itself. Though, we also do note that the muon mobility data was obtained on pressed powder pellets whereas the data presented here are for thin films of material and this may also impact the mobility of the ions. As such, we will have to wait on more information on the mobility of protons in thin films of melanin.

From the viewpoint of applications, given our suggestion of the presence of $H_5O_2^+$ suggests a change to the traditional focus in methods of enhancement of melanin conductivity. Currently, means of doping melanin has been achieved with metal ions exchanging protons in carboxyl and semiquinone hydroxyl groups [31]. However, the facilitated decomposition of $H_5O_2^+$ via thermal or chemical methods may become a fruitful alternative in some cases to help enhance device relevant thin films' conductivity behaviour.

Furthermore, the work indicates the need to determine the proton mobilities within melanin as a function of hydration.

## 5. Conclusions

The presented FTIR spectroscopy study armed with precise control of hydration level of material allowed for the first time to track the comproportionation reaction between monomer units of melanin polymer, which is manifested by a gradual decrease of carbonyl groups concentration (1724 cm$^{-1}$ band) with the increase of water content. Furthermore, analysis of the detected complex behavior of different water-based structures, suggests two types of interfacial water (3253 cm$^{-1}$ and 3473 cm$^{-1}$ bands). Finally, the band detected at 3600 cm$^{-1}$ has been assigned to a probable Zundel cation $H_5O_2^+$, where we suggest that it may act as a mobile proton trap in films of melanin.

**Author Contributions:** ZVB constructed the setup and carried out the hydration-controlled FTIR experiment. ZVB, SSZ and BPG performed spectra analysis. ABM synthesized and characterized the melanin sample. PAA and IOT made sample films and characterized their properties. KAM suggested the explanation of melanin hydration-dependent properties via Zundel cation formation, conceived and supervised the work. KAM, ZVB and ABM wrote the manuscript.

**Funding:** This work was supported by the Russian Science Foundation, grant 19-73-10154. A.B.M. is a Sêr Cymru II fellow and the results incorporated in this work is supported by the Welsh Government through the European Union's Horizon 2020 research and innovation program under the Marie Skłodowska-Curie grant agreement No 663830.

**Data Availability Statement:** The data that support the findings of this study are available from the corresponding author, K.M., upon reasonable request.

**Acknowledgments:** We would like to express our gratitude to Dr. Ivan Popov for his important and enduring explanations related to collective water molecules dynamics and proton transfer mechanisms in solids. We thank the Electron Paramagnetic Group at Cardiff University and Dr. Emma Richards in particular for the use of their EPR equipment. We also thank Dr. J.D. McGettrick, Swansea University, for helping obtain the XPS data. We thank Dr. Alexey Kuksin in particular for assistance with estimation of the sample films thickness.     In this section, you can acknowledge any support given which is not covered by the author contribution or funding sections. This may include administrative and technical support, or donations in kind (e.g., materials used for experiments).

# SUPPLEMENTARY MATERIALS

**Infrared spectroscopy of hydration-controlled eumelanin films suggests the presence of the Zundel cation**

Zakhar V. Bedran [1], Sergey S. Zhukov [1], Pavel A.Abramov [1], Ilya O. Tyurenkov [1], Boris P. Gorshunov [1], A. Bernardus Mostert [2],   and Konstantin A. Motovilov [1,*]

[1]      Center for Photonics and 2D Materials, Moscow Institute of Physics and Technology, Institute Lane 9, 141701, Dolgoprudny, Moscow region, Russia; info@mipt.ru
[2]      Department of Chemistry, Swansea University, Singleton Park, SA2 8PP, Swansea, Wales, UK; study@swansea.ac.uk

*       **Correspondence: k.a.motovilov@gmail.com**

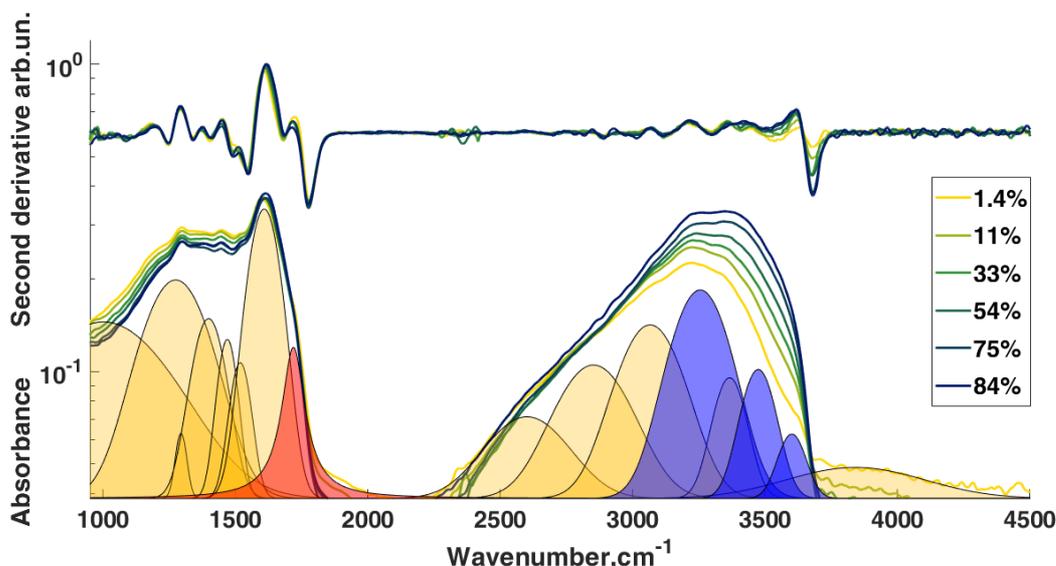

**Figure 1**. The solid green line is the measured infrared spectrum of the synthetic eumelanin thin film at 25ºC at relative humidity value (RH) 1.4% . The colored areas are the result of 1.4% RH spectra deconvolution into an appropriate set of well-known lines. The red peak captures the C=O vibrations, the three blue peaks are assigned to water vibrations and the remainder, yellow peaks referred to various excitations within the melanin system.  On the upper inset we show the second derivatives for all the data we obtained at the various hydration levels starting from 1.4% RH (yellow) and ending on 84% RH (blue).

**Table 1.** Positions of the observed peaks and their assignments.

| Second derivative peak position, cm$^{-1}$ | Line peak position, cm$^{-1}$ | Assignment |
|---|---|---|
| 1043 | 1000 ± 5 | δring indolequinone [1] |
| 1192 | 1273 ± 1 | ν(CO) + δ(CH) + δring [1] |
| 1291 | 1294 ± 1 | ν(CO) + δ(NH) + δ(CH) [1] |
| 1376 | 1398 ± 2 | δ(OH) + ν(ring) [1] |
| 1449 | 1468 ± 2 | ν(ring) + ν(CN) + δ(NH) + δ(OH) [1] |
| 1511 | 1517 ± 2 | ν(ring) + δ(NH) + δ(CH) [1] |
| 1613 | 1608 ± 2 | ν(ring) + δ(CH) [1] |
| 1712 | 1718 ± 1 | ν(C=O) in ketone or carboxylic acid [1] |
| 2590 | 2600 ± 2 | Carboxylic –OH stretch [1,2] |
| 2769 | - |  |
| 2849 | 2851 ± 2 | Aliphatic CH stretch [1,2] |
| 2932 | - |  |
| 3065 | 3065 ± 1 | Aromatic CH stretch [1,2] |
| 3209 | 3254 ± 3 | Stretching vibration of the Ice-like hydrogen-bonded water at melanin particles interface [Warren Brandt 2008] |
| 3360 | 3365 ± 1 | ν(NH) [1] |
| 3440 | 3473 ± 1 | Stretching vibration of the non-interface liquid-like water [3] |
| 3631 | 3600 ± 1 | $H_5O_2^+$ [4] |
| - | 3840 +\- 30 | Unknown overtone |

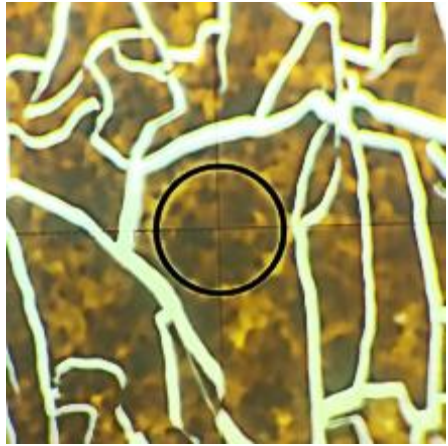

**Figure 2**. The microscope image of the measured synthetic melanin thin film. The black circle in the middle is the area exposed to IR radiation beam with a 50 $\mu$m diameter. The image was post-processed in PhotoFiltre Studio X to enhance contrast and to add central black circle.

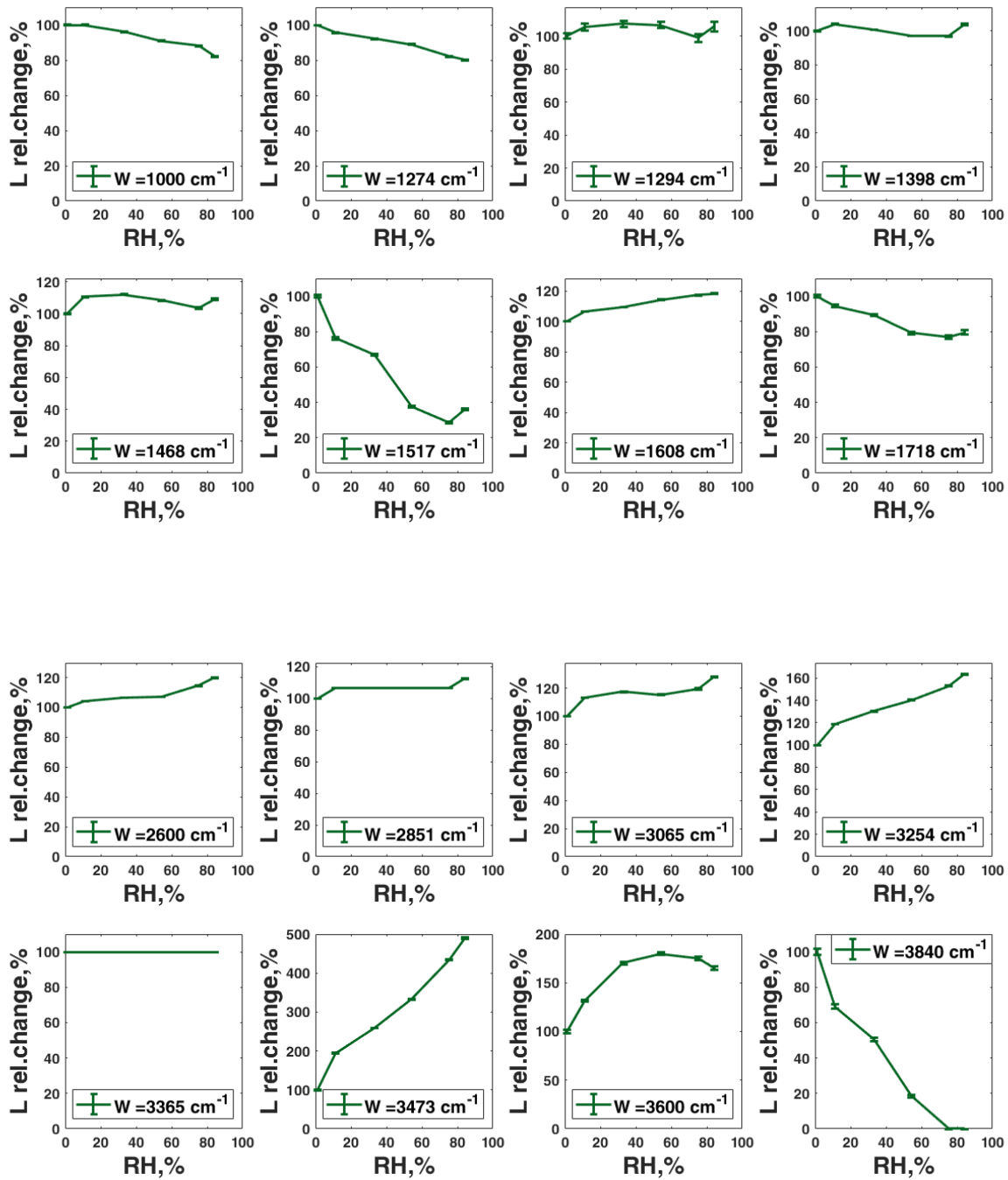

**Figure 3**. The observed evolution of the relative line strength L of the all detected modes with respect to relative humidity (RH). On the legend, W refers to the central wavenumber of the mode in cm$^{-1}$. The error bars were estimated from the L-M LSDM algorithm.

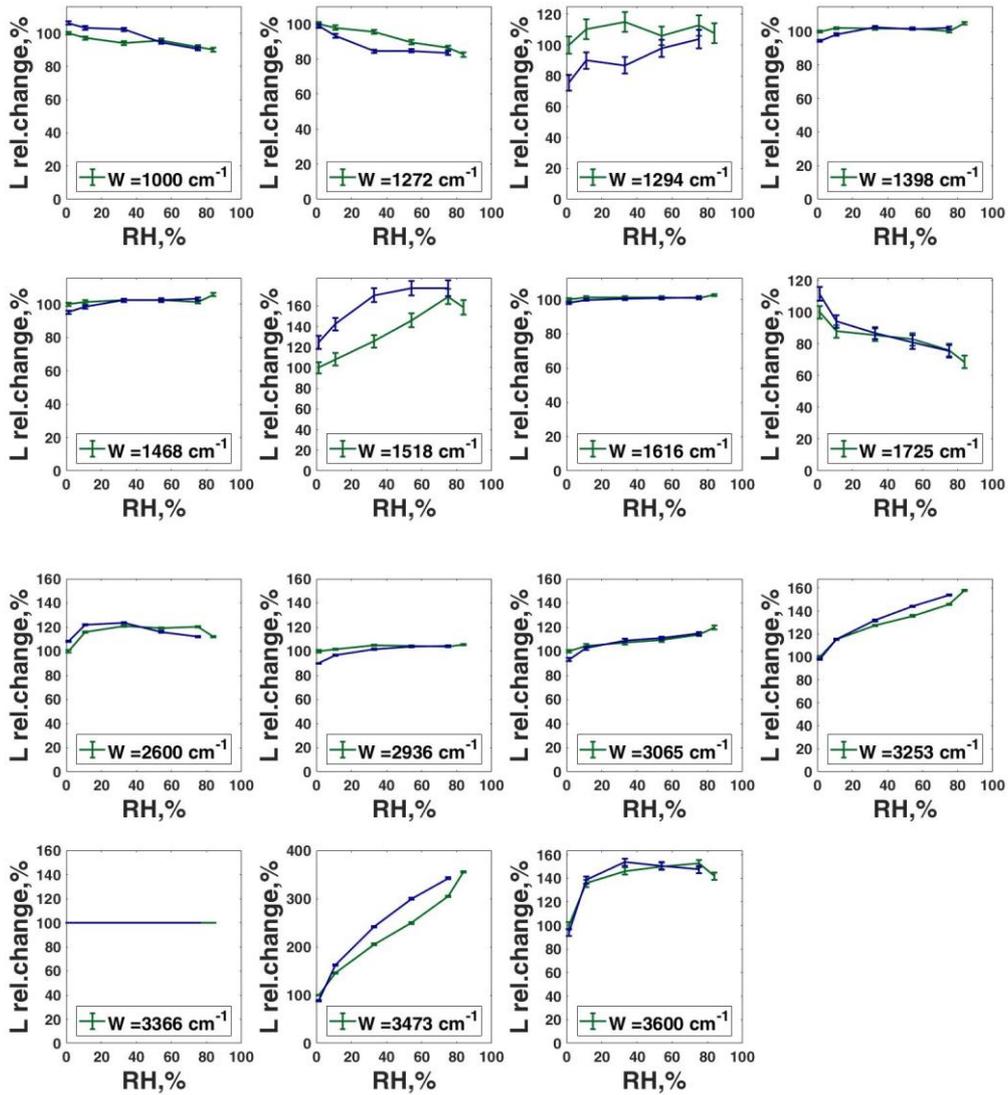

**Figure 4**. The observed evolution of the relative line strength with hydration of the all detected modes in the experiment with increase and decrease of the relative humidity (RH). On the legend, W refers to the central wavenumber of the mode in cm$^{-1}$. The error bars were estimated from the L-M LSDM algorithm. The green lines referred to the increase of RH during the measurements and the blue ones referred to the measurements with decrease of RH.

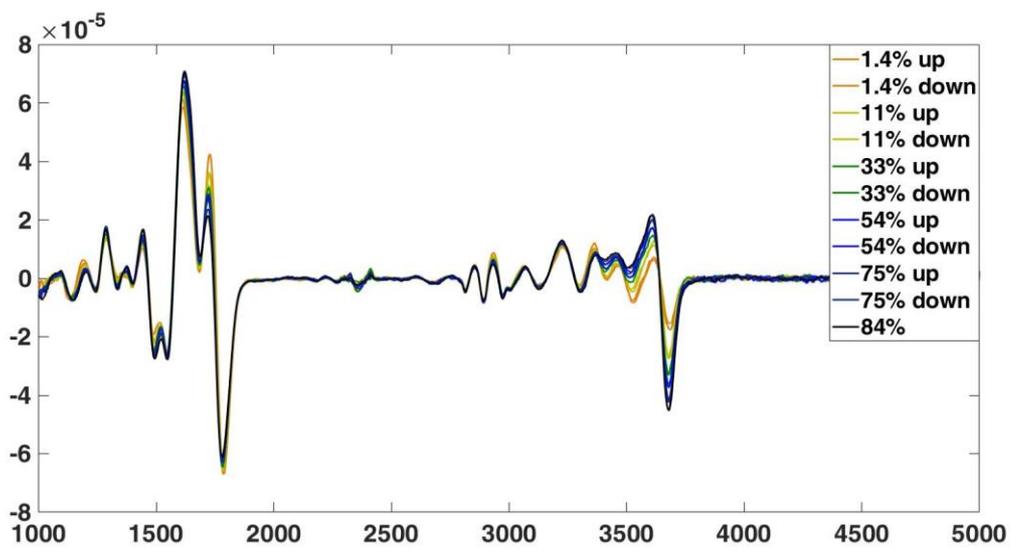

**Figure 5**. The values of the second derivative of absorbance for the experiment with the consequent increase and decrease of the relative humidity.

**Discussion: $H_5O_2^+$ is a proton hinderer in inorganic solid state proton conductors**

$^{31}$P, $^1$H, $^2$H and $^{17}$O NMR studies performed by Kolokolov et al [5] and Uchida et al [6,7] on variously hydrated solid 12-tunstophosphoric acid (TPA) $H_3PW_{12}O_{40} \times nH_2O$ gave the bright evidence of direct connection between type of dominating proton species in corresponding system, $H_3O^+$ or $H_5O_2^+$, and proton mobility. TPA represents a classic Keggin unit (central $PO_4$ tetrahedron is surrounded by 12 metal-oxygen octahedral $WO_6$) [8] able to form various self-organizing superstructures depending on environmental conditions. The unit charge (-3) is neutralized by acid surface protons. The most precise hydration control was performed in the study [5] on deuterated TPA. At n = 0.3 $^2$H NMR spectrum corresponds to static protons (deuterons of TPA) [9]. At n = 0.7 – 5.1 spectrum reflects the presence of two different species. The mobile one is associated with hydronium cations and hydrogen-bonded water [10,11]. The static component is responsible for surface protons of the unit. Gradual hydration leads to almost complete disappearance of static component at n = 5.1 and full activation of the mobile one. Further hydration of TPA leads to fundamental change in spectra important for the following discussion. At n = 5.2 authors [5] observe the complete disappearance of fully static components. Isotropic signal responsible for mobile proton species remain unchanged, but the new feature appears. It is associated with the fast anisotropic rotation of water around the C2 axis by 180° C. According to earlier quasi-elastic neutron scattering study it is responsible for internal two-site flips of water incorporated into $H_5O_2^+$ cation [12]. Further hydration up to n = 9 favors $H_5O_2^+$ component and diminishes those associated with the mobile proton species. Therefore $H_3O^+$ transforms into $H_5O_2^+$ in the system leading to remarkable decrease of general proton mobility, i.e. $H_5O_2^+$ is the main hinderer of proton mobility [5]. However, the story doesn't end here. Extra hydration changes TPA superstructure and brings mobile proton species again on the scene leading to increased proton mobility. The discussion devoted to diversity of mobile proton species in TPA at higher levels of hydration can be found elsewhere in the literature, but we return here to our main subject.

**ESI REFERENCES**